\documentclass{els}
\usepackage{epsfig}
\def\etal{{\it et al.}}
\begin{document}

\title{Weak Decays, CKM and CP Violation}
\author{Sheldon Stone}

\address{Physics Department, Syracuse University, Syracuse N. Y., 
USA\\E-mail: stone@phy.syr.edu}
\maketitle

\begin{center} {\bf Abstract} \end{center}
 I review several topics pertaining to Weak decays of $b$ and $c$ quarks, 
 including measurements of $|V_{cb}|$,
$|V_{ub}/V_{cb}|$, $f_{D_s}$ and $b\to s\gamma$.

\section{Introduction}

 Leptons and quarks, along with gluons, photons and gauge bosons are the 
fundamental objects in nature described by the Standard Model of electroweak 
interactions. Although the model has been successful at describing the 
interactions between these objects, many important questions remain.

\begin{tabbing}
\=$\bullet$Why are there so many fundamental constants?\\
\>$\bullet$What is the relationship of these constants to quark masses?\\
\>$\bullet$Are quarks and leptons really pointlike?\\
\>$\bullet$Is the Standard Model description correct, especially of CP
violation?\\
\>$\bullet$What is the connection between CP and matter-antimatter asymmetry?\\
\end{tabbing}

In weak interactions of quarks, we are interested in the couplings of quarks to each 
other and leptons, but have to deal with the ``brown muck''  of hadrons.
The basic weak $V-A$ structure has been verified with purely leptonic decays, 
for example, 
$\mu\to e\nu_e\nu_{\mu}$, $\tau\to e\nu_e\nu_{\tau}$. I do not have enough
 space to report on all interesting aspects of weak decays here, so I will 
report on a few, but miss others, even ones which I covered in
my presentation.

\subsection{The CKM Matrix and CP Violation}
\label{sec:Intro}

The physical point-like states of nature that have both strong and electroweak
interactions, the quarks, are mixtures of base states described by the
Cabibbo-Kobayashi-Maskawa matrix,\cite{ckm}
\begin{displaymath}
\left(\begin{array}{c}d'\\s'\\b'\\\end{array} \right) =
\left(\begin{array}{ccc} 
V_{ud} &  V_{us} & V_{ub} \\
V_{cd} &  V_{cs} & V_{cb} \\
V_{td} &  V_{ts} & V_{tb}  \end{array}\right)
\left(\begin{array}{c}d\\s\\b\\\end{array}\right)
\end{displaymath}
The unprimed states are the mass eigenstates, while the primed states denote
the weak eigenstates. There are nine complex CKM elements. These 18 
numbers can be reduced to four independent quantities by applying unitarity 
and the fact that the phases of the quark wave functions are 
arbitrary. 
These four remaining numbers are  fundamental constants of nature that 
need to be determined from experiment, like any other
fundamental constant such as $\alpha$ or $G$. In the Wolfenstein 
approximation the matrix is written as\cite{wolf}
\begin{displaymath}
V_{CKM} = \left(\begin{array}{ccc} 
1-\lambda^2/2 &  \lambda & A\lambda^3(\rho-i\eta(1-\lambda^2/2)) \\
-\lambda &  1-\lambda^2/2-i\eta A^2\lambda^4 & A\lambda^2(1+i\eta\lambda^2) \\
A\lambda^3(1-\rho-i\eta) &  -A\lambda^2& 1  
\end{array}\right).
\end{displaymath}
This expression is accurate to order $\lambda^3$ in the real part and
$\lambda^5$ in the imaginary part. It is necessary to express the matrix
to this order to have a complete formulation of the physics we wish to pursue.
The constants $\lambda$ and $A$ have been measured using semileptonic
$s$ and $b$ decays;\cite{virgin} $\lambda\approx 0.22$, and $A\approx 0.8$.

The phase $\eta$ allows for CP violation. 
CP violation thus far has only been seen in the neutral kaon 
system. If we can find CP violation in the $B$ system we could see
if the CKM  model works or perhaps discover new physics that 
goes  beyond the model, if it does not.

It is also of great interest to measure the magnitudes of each of the matrix
elements. Techniques used have included: $V_{ud}$ from 
$0^+\to 0^+$ nuclear $\beta$-decay, $V_{us}$ from $K\to\pi\ell\nu$ and hyperon
semileptonic decays, $V_{ub}$ from charmless semileponic $b$ decays, $V_{cd}$ 
from neutrino interactions and charm semileptonic
decay, $V_{cs}$ from direct $W^{\pm}$ decays at LEP II, $V_{cb}$ from charmed
semileptonic $b$ decays, $V_{td}$ from $B_d^o$ mixing, limits on $V_{ts}$ from
$B_s$ mixing, and limits on $V_{tb}$ from $t$ decays.
The measurements of $V_{cb}$ and $V_{ub}$ will be discussed here.

 \subsection{Measurement Of $|V_{cb}|$ Using $B\to D^*\ell\nu$}
Currently, the most favored
technique is to measure the decay rate of $B\to D^{*}\ell^-\bar{\nu}$ at the
kinematic point where the $D^{*+}$ is at rest in the $B$ rest frame (this is 
often referred to as maximum $q^2$ or $\omega =1$). Here, according to Heavy
Quark Effective Theory, the theoretical 
uncertainties are at a minimum.

\begin{figure}
\vspace{-19mm}
\centerline{\epsfig{figure=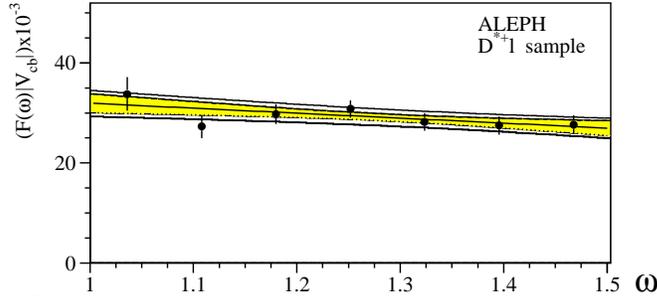,width=4.2in}}
\vspace{-50mm}
\caption{\label{aleph_vcb} $\overline{B}^o\to D^{+}\ell^-\bar{\nu}$ from ALEPH. 
The data have been
fit to a functional form suggested by Caprini \etal ~The abcissa gives the
value of the product $|F(1)*V_{cb}|^2$.}
\end{figure}

There are results from several groups using this technique for the decay sequence 
$D^{*+}\to \pi^+ D^o$; $D^o\to
K^-\pi^+$, or similar decays of the $D^{*o}$. The ALEPH results\cite{aleph_pi} are shown in
Fig.~\ref{aleph_vcb}.

In a recent analysis, DELPHI detects  only the slow $\pi^+$ from the $D^{*+}$ 
decay and does not 
explicitly reconstruct the $D^o$ decay.\cite{delphi_pi} 
Table~\ref{tab:Vcb} summaries determinations of $|V_{cb}|$; here,
the first error is statistical, the second systematic and the third, an estimate 
of the theoretical accuracy in predicting the form-factor $F(\omega=1)=0.91\pm
0.003$.\cite{formfactor}
Currently, DELPHI has the smallest error, 
however, CLEO has only used 1/6 of their current data. The quoted average $|V_{cb}|=0.0381\pm 0.0021$
combines the averaged statistical and systematic errors with the theoretical 
error in quadrature and
takes into account the common systematic errors, such as the $D^*$ branching
ratios.

\begin{table}[th]
\vspace{-2mm}
\begin{center}
\caption{Modern Determinations of $|V_{cb}|$ using 
$B\to D^*\ell^-\overline{\nu}$ decays
at $\omega = 1$ \label{tab:Vcb}}
\vspace*{2mm}
\begin{tabular}{lc}\hline\hline
Experiment & $V_{cb}$ $(\times 10^{-3})$\\\hline
ALEPH\cite{aleph_pi} & $34.4\pm 1.6 \pm 2.3 \pm 1.4$ \\
DELPHI\cite{delphi_pi} & $41.2\pm 1.5 \pm 1.8 \pm 1.4$ \\
OPAL\cite{opal_pi} & $36.0\pm 2.1 \pm 2.1 \pm 1.2$ \\
CLEO\cite{cleo_pi} & $39.4\pm 2.1 \pm 2.0 \pm 1.4$ \\\hline
Average & $38.1\pm 2.1$\\
 \hline\hline
\end{tabular}
\end{center}
\end{table}

There are other ways of determining $V_{cb}$. One new method based on QCD sum rules uses the 
operator product expansion and the heavy quark expansion, in terms of  the parameters $\alpha_s (m_b)$, $\overline{\Lambda}$, and the matrix elements $\lambda_1$ and $\lambda_2$. 
The latter quantities arise from the differences
\begin{displaymath}
m_B-m_b=\overline{\Lambda}-{{\lambda_1+3\lambda_2}\over{2m_b}}~~~
m_B^*-m_b=\overline{\Lambda}-{{\lambda_1+-\lambda_2}\over{2m_b}}~~.
\end{displaymath}
 The $B^*-B$ mass difference determines $\lambda_2 = 0.12$ GeV$^2$.
The total semileptonic decay width is then related to above parameters as
\begin{eqnarray*}
&&\Gamma_{sl}={{G_F^2\left|V_{cb}\right|^2m_B^5}\over
{192\pi^3}}0.369\times\nonumber\\
&&\left[1-1.54{\alpha_s \over \pi}-1.65{\overline{\Lambda}\over m_B}\left(
1-.087{\alpha_s\over\pi}\right)-0.95{\overline{\Lambda}^2\over m_B^2}
-3.18{\lambda_1\over m_B^2}+0.02{\lambda_2\over m_B^2}\right]\nonumber\\
\end{eqnarray*}

CLEO has measured the semileptonic branching ratio using lepton tags as 
(10.49$\pm$0.17$\pm$0.43)\% and using the world average lifetime for an equal 
mixture of $B^o$ and $B^-$ mesons of  1.613$\pm$0.020 ps, CLEO finds
$\Gamma_{sl} = 65.0\pm 3.0$ ns$^{-1}$. (Note that LEP has a somewhat larger 
value of  68.6$\pm$1.6 ns$^{-1}$.)

\begin{figure}[b]
\vspace{-5mm}
\centerline{\epsfig{figure=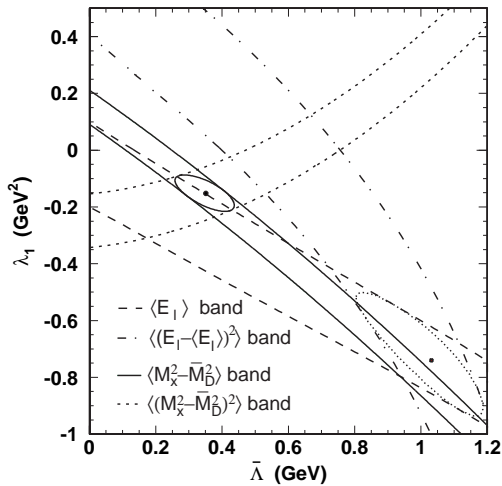,width=2.1in}}
\vspace{5mm}
\caption{\label{moments}Bands in $\overline{\Lambda}-\lambda_1$ space found
by CLEO in analyzing first and second moments of hadronic mass squared and
lepton energy. The intersections of the two moments for each set determines
the two parameters. The one standard deviation error ellipses are shown.}
\end{figure}

CLEO then attempts to measure the remaining unknown parameters $\lambda_1$ 
and $\overline{\Lambda}$ by using moments of the either the hadronic mass or 
the lepton energy.\cite{moment} The results are shown in Fig.~\ref{moments}. Here the 
measurements are
shown as bands reflecting the experimental errors. Unfortunately, this
preliminary CLEO result shows a contradiction. The overlap of the mass moment 
bands gives different values than the lepton energy moments! The mass
moments are theoretically favored and give the values $\lambda_1$=
(0.13$\pm$0.01$\pm$0.06) GeV$^2$, and  $\overline{\Lambda}$ = 
(0.33$\pm$0.02$\pm$0.08) GeV. The discrepancy between the two methods is
serious. It either means that there is something wrong with the CLEO analysis
 or there is something wrong in the theory. If the latter is true it would shed
doubt on the method used by the LEP experiments to extract a value of $|V_{ub}|$
using the same theoretical framework.

\subsection{Measurement Of $|V_{ub}|$}
Another important CKM element that can be measured using semileptonic decays is
$V_{ub}$. The first measurement of $V_{ub}$ done by CLEO and subsequently
confirmed by ARGUS, used only leptons which were more energetic than those that
could come from $b\to c\ell^- \bar{\nu}$ decays.\cite{first_vub} These  
``endpoint leptons'' can occur, $b\to c$ background free, at the
$\Upsilon (4S)$ because the $B$'s are almost at rest. Unfortunately, there is
only  a small fraction of the $b\to u \ell^-\bar{\nu}$ lepton spectrum that
can be seen this way, leading to model dependent errors.

\begin{figure}[bt]
\vspace{-9mm}
\centerline{\epsfig{figure=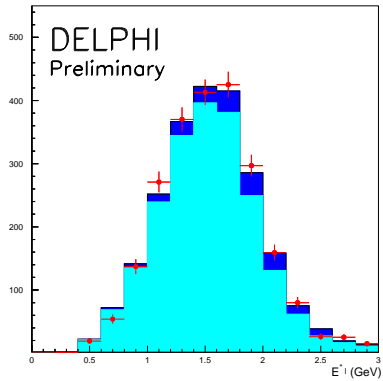,width=2.in}}
\vspace{-4mm}
\caption{\label{delphi_vub}The lepton energy distribution in the $B$ rest
frame from DELPHI. The data have been enriched in $b \to u$ events, and the 
mass of the recoiling hadronic system is required to be below 1.6 GeV. The 
points indicate data, the light shaded region, the fitted background and the 
dark shaded region, the fitted $b \to u \ell \nu$ signal. }
\end{figure}

ALEPH\cite{aleph_vub} L3\cite{L3_vub}
and DELPHI\cite{delphi_vub} try to isolate a class of events where the hadron 
system associated
with the lepton is enriched in $b\to u$ and thus depleted in $b\to c$.     
They define a likelihood that hadron tracks come from $b$ decay by using a large 
number of variables including, vertex information, transverse momentum, not 
being a kaon. Then they require the hadronic mass to be less than 1.6 GeV, which 
greatly reduces $b\to c$, since a completely reconstructed $b\to c$ decay has a 
mass greater than that of the $D$ (1.83 GeV). They then examine the lepton 
energy distribution for this set of events, shown in Fig.~\ref{delphi_vub} for
DELPHI.

\begin{figure}[htb]
\centerline{\epsfig{figure=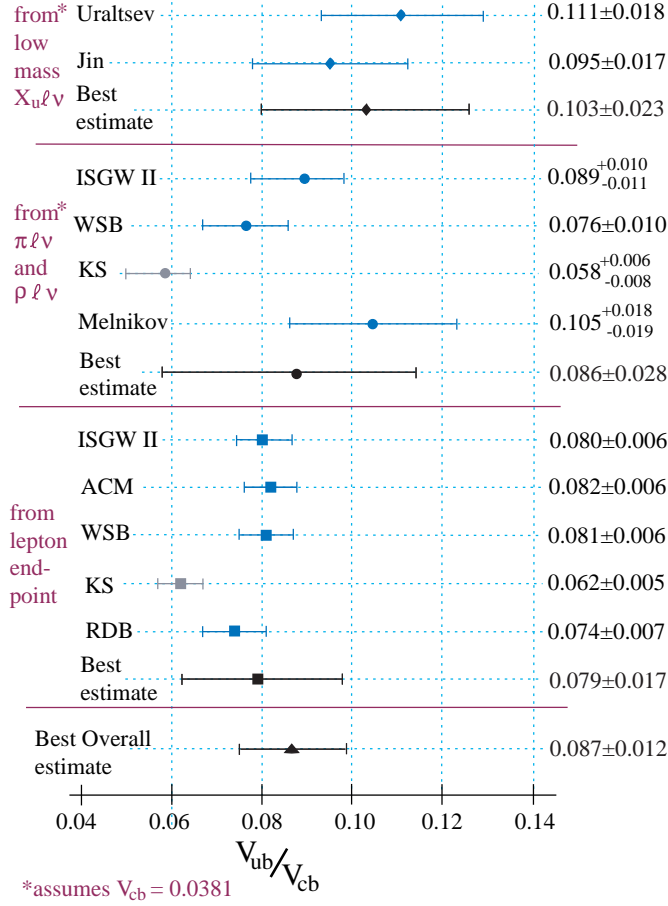,width=3.6in}}
\caption{\label{vub}Measurements of $|V_{ub}/V_{cb}|$ using different
techniques and theoretical models. (The KS model has been ruled 
out.) }
\end{figure}

 I have averaged all three LEP results and show them in 
Fig.~\ref{vub} without any theoretical error, which is estimated
at $\pm$8\% by Uraltsev.\cite{vub_thy_inc} However, another calculation 
using the same type of model by Jin\cite{jin} gives a $\pm$14\% lower value, with a 
quoted error of $\pm$10\%. 
 
My best estimate of $\left|V_{ub}/V_{cb}\right|$ 
using this technique includes a $\pm$14\% theoretical error added in quadrature
with a common systematic 
error of $\pm$14\%, since the Monte Carlo calculations at LEP are known to be 
strongly correlated. 

Also shown in Fig.~\ref{vub} are results from CLEO using the measured the decay 
rates for the exclusive final states $\pi\ell\nu$ and 
$\rho\ell\nu$,\cite{cleo_pirho} and results from endpoint leptons, dominated 
by CLEO II.\cite{cleo_vub} Several theoretical
models are used.\cite{vub_thy} From the exclusive results, the model of
Korner and 
Schuler (KS) is ruled out by the measured ratio of $\rho/\pi$. This model deviated 
the most from the others used to get values of $|V_{ub}|$ from endpoint leptons. 
Thus the main use of the exclusive final states has been to restrict the 
models. The endpoint lepton results are statistically the most precise. 
Assigning a model dependent error is quite difficult. I somewhat arbitrarily 
have assigned a $\pm$14\% irreducible systematic error to these models and used the 
average among them to derive a value.
My best overall estimate is that $|V_{ub}/V_{cb}|=0.087\pm 0.012$.

This estimate must be treated as highly suspect. The value and error depends on
uncertain theoretical estimates. We can use this estimate, along with other
measurements. To get some idea of what the values of $\rho$ and $\eta$ are.

\begin{figure}[htbp]
\centerline{\epsfig{figure=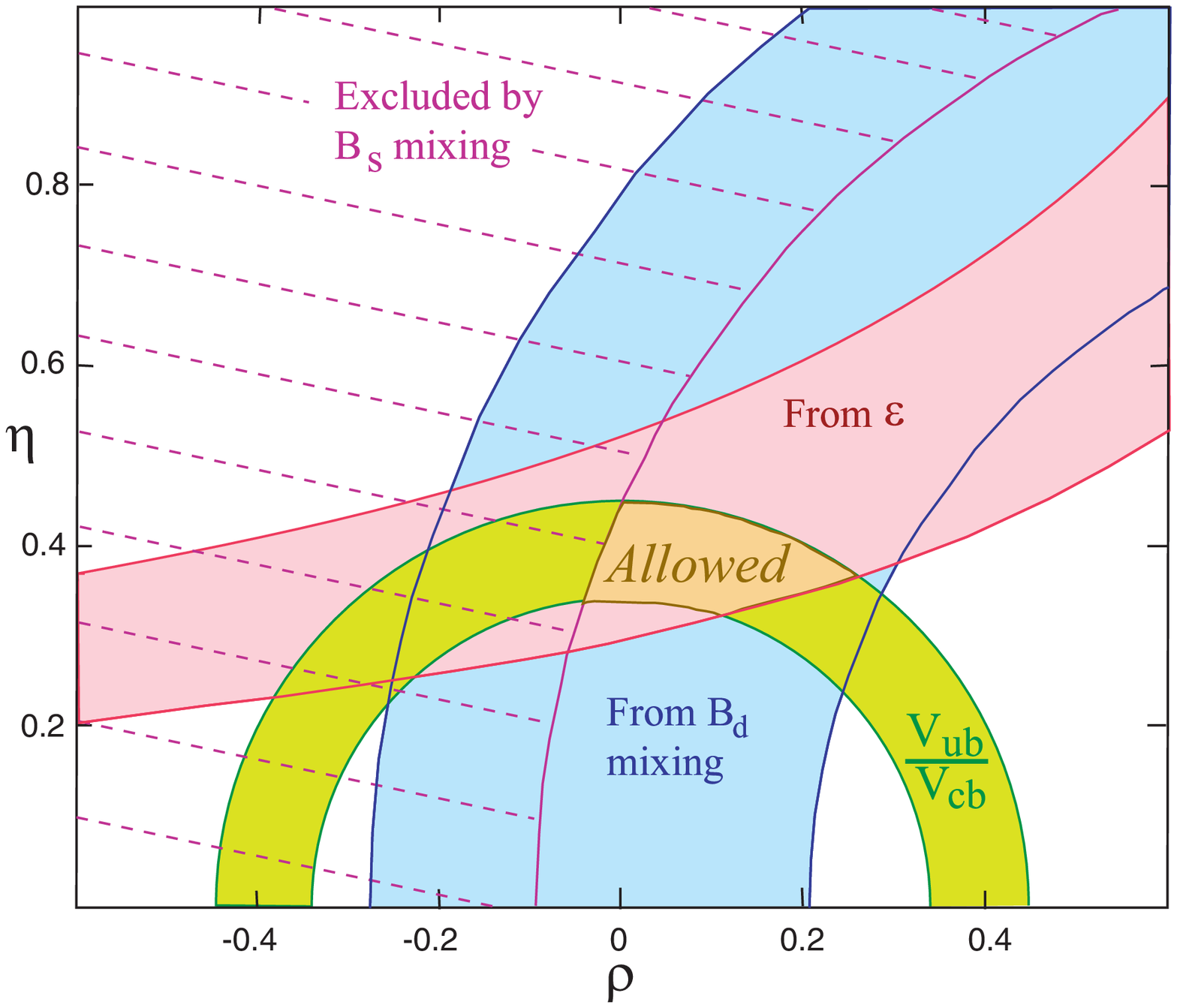,height=2.5in}}
\vspace{-.4cm}
\caption{\label{ckm_tri_6}The regions in $\rho-\eta$ space (shaded) consistent
with measurements of CP violation in $K_L^o$ decay ($\epsilon$), $V_{ub}/V_{cb}$
in semileptonic $B$ decay, $B_d^o$ mixing, and the excluded region from
limits on $B_s^o$ mixing. The allowed region is defined by the overlap of
the 3 permitted areas, and is where the apex of the  CKM triangle  sits.}
\end{figure}

There is a constraint on $\rho$ and $\eta$ given by the $K_L^o$ CP violation 
measurement ($\epsilon$), given by\cite{buras}
\begin{displaymath}
\eta\left[(1-\rho)A^2(1.4\pm 0.2)+0.35\right]A^2{B_K \over 0.75}=(0.30\pm 0.06),
\end{displaymath}
where the errors arise mostly from uncertainties on $|V_{cb}|$ and $B_K$. 
Here $B_K$
is taken as 0.75$\pm$0.15 according to Buras.\cite{Buras_bk} The constraints on 
$\rho$ versus $\eta$ from the $|V_{ub}/V_{cb}|$ determination, 
$\epsilon$ and $B$ 
mixing are shown in Fig.~\ref{ckm_tri_6}. The bands represent $\pm1\sigma$ errors,
for the measurements and a 95\% confidence level upper limit on $B_s$ mixing.
 The width of the $B_d$ mixing band 
is caused mainly by 
the uncertainty on $f_B$, taken here as $240> f_B > 160$ MeV. Other parameters
include $|V_{cb}|=0.381\pm 0.0021$, $|V_{ub}/V_{cb}| = 0.087\pm 0.012$, limit on 
$\Delta M_s > 12.4$ ps$^{-1}$, and the ratio
$f_{B_s}\sqrt{B_{B_s}}/f_{B_d}\sqrt{B_{B_d}}\leq 1.25$.\cite{Jonnew}

\section{The decays $B^-\to\ell^-\overline{\nu}$ and $D_s^+\to\mu^+\nu$}

\begin{figure}[htb]
\vspace{-.2cm}
\centerline{\epsfig{figure=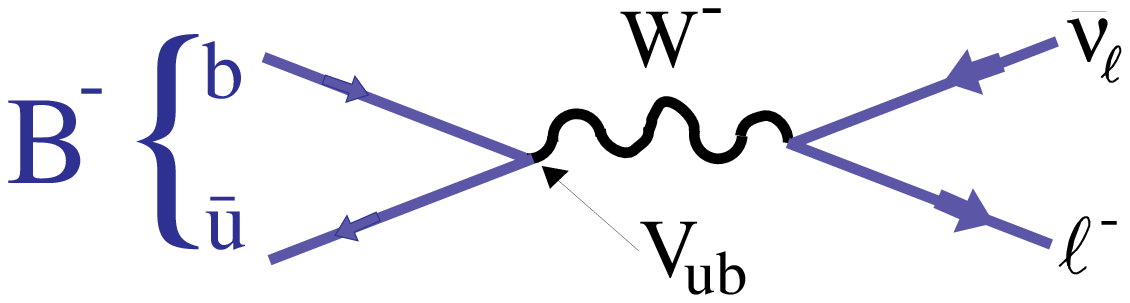,height=1.2in}}
\vspace{-1.8cm}
\caption{\label{btolnu} Diagram for a $B^-\to \ell^-\overline{\nu}$ decay.}
\end{figure}

This reaction proceeds via the annihilation of the $b$ quark with the
$\overline{u}$ into a virtual $W^-$ which materializes as
$\ell^-\overline{\nu}$ pair as illustrated in Fig.~\ref{btolnu}. The decay rate for this process can be written
as
\begin{displaymath}
\Gamma(B^-\to \ell^- \overline{\nu})={{G_F^2}\over 8\pi}f_{B}^2m_{\ell}^2M_{B}
\left(1-{m_{\ell}^2\over M_{B}^2}\right)^2 \left|V_{ub}\right|^2~~~,
\label{eq:equ_rate}
\end{displaymath}
where $f_B$ is the so called ``decay constant," a parameter that can be
calculated theoretically or determined by measuring the decay rate. This
formula is the same for all pseudoscalar mesons using the appropriate CKM
matrix element and decay constant.

Knowledge of $f_B$ is important because it is used to determine constraints on
CKM matrix elements from measurements of neutral $B$ mixing. Since the decay is
helicity suppressed, the heavier the lepton the larger the expected rate. Thus
looking for the $\tau^-\overline{\nu}$ has its advantages. The big disadvantage
is that there are least two missing neutrinos in the final state. 
The most stringent limit has been set by L3 of $<5.7\times
10^{-4}$ at 90\% confidence level, using a missing energy technique.\cite{L3taunu} This is
still one order of magnitude higher than what is expected. Other limits
are poorer.\cite{othertaunu}

Since $f_B$ is so difficult to measure, models, especially lattice gauge
models,
are used.\cite{Bernard} However, it is prudent to test these models. 
$D_s^+\to\mu^+\nu$ can be used; it is Cabibbo favored and the predicted
branching ratio is close to 1\%.

\begin{figure}[b]
\vspace{-0.8cm}
\centerline{\epsfig{figure=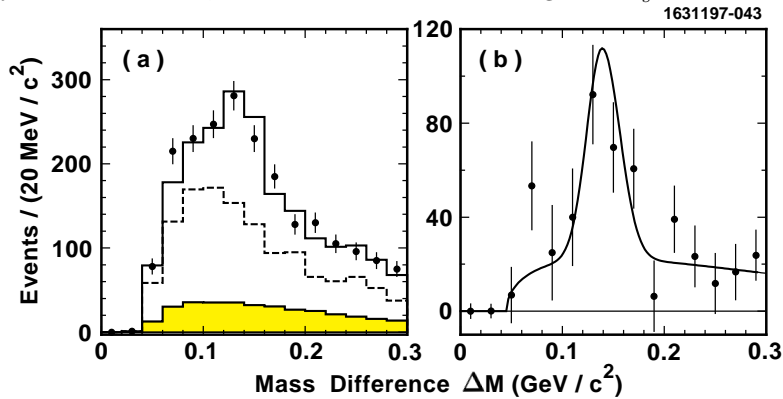,width=4.1in}}
\caption{\label{Fdata} (a) The $\Delta M$ mass difference distribution for 
$D_s^{*+}$
candidates for both the muon data (solid points), the electron data (dashed
histogram) and the excess of muon fakes over electron fakes (shaded). The histogram is
the result of the fit described in the text.
(b) The $\Delta M$ mass difference distribution for $D_s^{*+}$
candidates with electrons and excess muon fakes subtracted. The curve is a fit
to the signal shape described in the text.}
\end{figure}

CLEO has made the highest statistics measurement to date of
 ${\cal B}(D^+_s\to \mu^+\nu)$, by searching for the decay sequence
$D_s^{*+}\to\gamma D_s^+$, $D^+_s\to \mu^+\nu$. Since the decay $D_s\to e\nu$
is suppressed by four orders of magnitude due to helicity, they use this mode
to measure the physics backgrounds due to real muons. Then they need correct
only for differences in muon and electron efficiencies and fake rates.
They use missing energy
and momentum to define the $\nu$ direction.  The mass difference
$\Delta M$ is calculated as difference in $D_s^*$ and $D_s$ invariant
mass.  
The $\Delta M$ distributions for the muon and electron data and the calculated
effective excess of muon fakes over electron fakes are shown in
Fig.~\ref{Fdata}(a). The histogram is the result of a $\chi^2$ fit of the muon
spectrum to the sum of three contributions: the signal, the scaled electrons,
and the excess of muon over electron fakes. Here, the sizes of the electron and
fake contributions are fixed and only the signal normalization is allowed to
vary. The signal consists of two components, whose
relative normalization is fixed. These two components are the decay
$D_s^{*+}\to\gamma D_s^+$, $D_s^+\to\mu^+\nu$ and the direct decay
$D_s^+\to\mu\nu$ and $D^+\to\mu^+\nu$ combined with a random photon.

CLEO finds a signal of
182$\pm$22 events in the peak which are attributed to the process
$D_s^{*+}\to\gamma D_s^+$, $D_s^+\to\mu^+\nu$. They also find 250$\pm$38 events
in the flat part of the distribution corresponding to $D_s^+\to\mu^+\nu$ or
$D^+\to\mu^+\nu$ decays coupled with a random photon. The contribution  of a
real $D^+\to\mu^+\nu$ decay with random photons is not entirely negligible
since the $D^{*+}\to\gamma D^+$ branching ratio does not enter. The $D^+$
fraction is estimated to be about (18$\pm$8)\% relative to the total
$D_s^+\to\mu^+\nu$ plus random photon contribution.

 Several other groups have made measurements.
 The results are shown in
Table~\ref{tab:dsmunu}. I have changed the values of $f_{D_s}$ according
to the updated PDG $D_s$ decay branching fractions for the normalization
modes,\cite{PDG} and
have corrected the old CLEO result by using the new fake rates determined in
their updated analysis. In addition, there are new results using the $D_s^+\to \tau^+\nu$
decay from the L3 collaboration\cite{L3taunu} of ($309\pm 58\pm 33 \pm 38$) MeV, and 
($330 \pm 95$) MeV from the DELPHI collaboration.\cite{othertaunu} 
The world average value for $f_{D_s}$ is ($255\pm 21\pm 28$) MeV, where the
common systematic error is due the error on the absolute branching ratio for
$D_s^+\to \phi\pi^+$. These numbers are consistent with C. Bernard's world
average for lattice theories of (221$\pm$25) MeV.\cite{Bernard}

\begin{table}[hbt]
\vspace{-2mm} 
\caption{Measured values of $f_{D_s}$ from experimental
values of $\Gamma(D_s^+\to\mu^+\nu)$\label{tab:dsmunu}}
\begin{center} \footnotesize 
\begin{tabular}{|lccc|}\hline
Collaboration & Observed & Published $f_{D_s}$  &
	Corrected $f_{D_s}$ \\
& Events & value (MeV) & value (MeV) \\ \hline
CLEO (old) \cite{cleo} & 39$\pm$8 & $344\pm 37 \pm 52 \pm 42$ &
	 $282 \pm 30 \pm 43 \pm 34$  \\ 
WA75 \cite{Bullshit} & 6 & $232 \pm 45 \pm 20 \pm 48$ &
$ 213 \pm 41 \pm 18 \pm 26$\\
BES \cite{Bes} & 3 & $430 ^{+150}_{-130} \pm 40$ & Same\\ 
E653 \cite{E653} & $23.2\pm 6.0 ^{+1.0}_{-0.9}$ & $194\pm 35\pm 20\pm 14$ 
& $200\pm 35\pm 20\pm 26$\\
CLEO \cite{chanda} & 182$\pm$22 & - &
	 $280 \pm 19 \pm 28 \pm 34$  \\ 
\hline
\end{tabular}
 \end{center} \end{table}

\section{Rare Decays as Probes beyond the Standard Model}

Rare decays have loops in the decay diagrams so they are sensitive to high mass 
gauge bosons and fermions. Thus, they are sensitive to new physics.
However, it must be kept in mind that any new effect 
must be consistent with already measured phenomena such as $B_d^o$ mixing and 
$b\to s\gamma$.

\begin{figure}[htb]
\vspace{-.6cm}
\centerline{\epsfig{figure=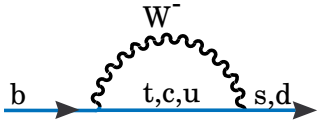,height=0.7in}}
\caption{\label{loop}Loop or ``Penguin" diagram for a $b\to s$ or $b\to d$
transition.}
\vspace{-1mm}
\end{figure}

These processes are often called
``Penguin" processes, for unscientific reasons. A Feynman
loop diagram is shown in Fig.~\ref{loop} that describes the transition of a $b$
quark into a charged -1/3 $s$ or $d$ quark, which is effectively a
neutral current transition. The dominant charged current
decays change the $b$ quark into a charged +2/3 quark, either $c$ or $u$.

The intermediate quark inside the loop can be any charge +2/3 quark. The relative
size of the different contributions arises from different quark masses
and CKM elements. In terms of the Cabibbo angle ($\lambda$=0.22), we have
for $t$:$c$:$u$ - $\lambda^2$:$\lambda^2$:$\lambda^4$. The mass dependence 
favors the $t$ loop, but the amplitude for $c$ processes can be 
quite large $\approx$30\%. Moreover, as
pointed out by Bander, Silverman and Soni,\cite{BSS} interference can occur
between $t$, $c$ and $u$ diagrams and lead to CP violation. In the standard model it is
not expected to occur when $b\to s$, due to the lack of a CKM phase difference,
but could occur when $b \to d$. In any case, it is always worth looking for
this effect; all that needs to be done, for example, is to compare the number
of $K^{*-}\gamma$ events with the number of $K^{*+}\gamma$ events.

There are 
other possibilities for physics beyond the standard model to appear. 
For example, the $W^-$
in the loop can be replaced by some other charged object such as a Higgs; it is
also possible for a new object to replace the $t$.

\subsection{$b \to s\gamma$}

This process occurs when any of the charged particles in Fig.~\ref{loop} emits
a photon.
CLEO first measured the inclusive rate\cite{oldbsg} as well as the exclusive rate into
$K^*(890)\gamma$.\cite{fbsg} There is an 
 updated CLEO measurement\cite{CLEObsg}
 using 1.5 times
the original data sample and a new measurement from
ALEPH.\cite{ALEPHbsg}

To remove background CLEO used two techniques originally, one based on ``event
shapes" and the other on summing exclusively reconstructed $B$ samples.
 CLEO uses eight
different shape variables,\cite{oldbsg} and defines a
variable $r$ using a neural network to distinguish signal from  background. The
idea of the $B$ reconstruction analysis is to find the inclusive branching
ratio by summing over exclusive modes. The allowed hadronic system is comprised
of either a $K_s\to\pi^+\pi^-$ candidate or a $K^{\mp}$ combined with 1-4
pions, only one of which can be neutral. The restriction on the number and kind
of pions maximizes efficiency while minimizing background. It does however lead
to a model dependent error. Then both analysis techniques are combined.
Currently, most of the statistical power of the
analysis ($\sim$80\%) comes from summing over the exclusive modes.

\begin{figure}[bh]
\centerline{\epsfig{figure=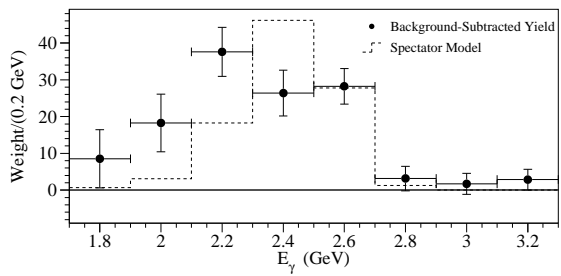,height=1.4in}}
\caption{\label{bsg_van}The background subtracted photon energy spectrum from
CLEO. The dashed curve is a spectator model prediction from Ali and Greub.}
\end{figure}

Fig.~\ref{bsg_van} shows the photon energy spectrum of the inclusive signal,
compared with the model of Ali and Greub.\cite{Ali} A fit to the model 
over the photon energy range from 2.1 to 2.7 GeV/c gives the branching 
ratio result shown in Table~\ref{btosgresults}, where the first error is
statistical and the second systematic. 

\begin{table}[hbt]
\vspace{-5mm} 
\caption{Experimental results for $b\to s\gamma$\label{btosgresults}}
\begin{center} \footnotesize 
\begin{tabular}{|lc|} \hline 
Sample & branching ratio\\\hline
CLEO & $(3.15\pm 0.35\pm 0.41)\times 10^{-4}$\\
ALEPH & $(3.11\pm 0.80\pm 0.72)\times 10^{-4}$ \\
Average & $(3.14\pm 0.48)\times 10^{-4}$\\
Theory\cite{bsgthy}&$(3.28\pm 0.30)\times 10^{-4}$\\
 \hline \end{tabular} \end{center} \end{table}

ALEPH reduces the backgrounds by weighting candidate decay
tracks in a $b \to s\gamma$ event by a combination of their momentum, impact
parameter with respect to the main vertex and rapidity with respect to the
$b$-hadron direction.\cite{ALEPHbsg} There result is shown in 
Table~\ref{btosgresults}. The world average
value experimental value is also given, as well as the theoretical prediction.


The consistency with standard model expectation has ruled out many models.
Hewett has  given a good review of the many minimal supergravity models which are
excluded by the data.\cite{Hewett}

Triple gauge boson couplings are of great interest in checking the standard
model. If there were an anomalous $WW\gamma$ coupling it would serve to change
the standard model rate. $p\overline{p}$ collider experiments have also
published results limiting such couplings.\cite{D0} In a two-dimensional space
defined by $\Delta\kappa$ and  $\lambda$, the D0 constraint appears as a tilted
ellipse and and the $b\to s\gamma$ as  nearly vertical bands. In the standard
model both parameters are zero.

\section*{Acknowledgments}
I thank Marina Artuso, B. Kayser, R. Peccei, Jon Rosner, and Tomasz Skwarnicki 
for interesting discussions.
This work was supported by the U. S. National Science Foundation.

\end{document}